\documentclass[11pt,a4paper]{article}
\usepackage{amsfonts}
\usepackage[dvipsnames,usenames]{color}
\usepackage[english]{babel}
\usepackage{ucs}
\usepackage[utf8x]{inputenc}
\usepackage{lmodern}
\usepackage[T1]{fontenc}
\usepackage{graphicx,amssymb,amsmath,bbm}
\usepackage{slashed,epstopdf}
\usepackage{tikz}
\usepackage{mathtools} 
\usepackage[indentafter,nobottomtitles]{titlesec}
\usepackage{titletoc}
\usepackage{fancyhdr}

\def\ii{\'{\i}}
\newcommand{\eref}[1]{(\ref{#1})}

\newcommand{\diagentry}[1]{\mathmakebox[1.8em]{#1}}
\newcommand{\xddots}{%
  \raise 4pt \hbox {.}
  \mkern 6mu
  \raise 1pt \hbox {.}
  \mkern 6mu
  \raise -2pt \hbox {.}
}

\usepackage{wasysym}

\newcommand{\red}[1]{{\color{red}#1}}
\newcommand{\darkerred}[1]{{\color{BrickRed}#1}}
\newcommand{\lighterred}[1]{{\color{OrangeRed}#1}}
\newcommand{\orange}[1]{{\color{BurntOrange}#1}}

\newcommand{\darkergreen}[1]{{\color{OliveGreen}#1}}

\newcommand{\blue}[1]{{\color{blue}#1}}
\newcommand{\darkerblue}[1]{{\color{NavyBlue}#1}}

\newcommand{\lighterblue}[1]{{\color{cyan}#1}}

\newcommand{\purple}[1]{{\color{Magenta}#1}}
\newcommand{\darkerpurple}[1]{{\color{Violet}#1}}
\newcommand{\beq}{\begin{eqnarray}}
\newcommand{\eeq}{\end{eqnarray}}
\newcommand{\R}{\mathbb{R}}
\newcommand{\Z}{\mathbb{Z}}

\def\Dsl{\,\raise.15ex\hbox{/}\mkern-12.5mu D}
\def\half{\textstyle{\frac12}}

\usepackage [english]{babel}
\usepackage{amsfonts}
\usepackage{amssymb}
\usepackage{epsf}
\usepackage{latexsym,amssymb,euscript}
\usepackage{graphicx}
\usepackage{amsmath}
\input epsf

\title{\bf{The critical transition of Coulomb impurities in gapped graphene}}
\author{\Large{M. Asorey and  A. Santagata}\\
\\
 Centro de Astropart\ii culas y F\ii sica de Altas Energ\ii as\\
  Departamento de F\'{\i}sica Te\'orica\\
 Universidad de
 Zaragoza, Zaragoza, Spain.}

\begin{document}
\def\today{}

\maketitle

\begin{abstract}

The effect of  supercritical  charge impurities in graphene is very similar 
to the supercritical  atomic collapses  in QED for $Z > 137$, but with a much lower critical charge. 
In this sense graphene can be considered as a natural 
testing ground for the analysis of quantum field theory vacuum instabilities.
We analyze the quantum transition from subcritical to supercritical charge regimes in gapped graphene
in a common framework that preserves unitarity for any value of charge  impurities. 
In the supercritical regime it is possible to introduce  boundary conditions which control the singular
behavior at the impurity. We show that for subcritical charges there are also 
non-trivial boundary conditions which are  similar to those that appear in  QED  
for nuclei in the intermediate regime $118<Z<137$.
We analyze the behavior of  the energy levels associated to the different boundary conditions.
In particular, we point out the existence of new bound states in the subcritical regime which include
a negative energy bound state in the attractive Coulomb regime.
A remarkable property is the continuity of the energy spectral flow  under variation of  the impurity  charge 
 even when jumping across the critical charge  transition. We also remark that the energy levels of Hydrogenoid 
bound states at critical values of  charge impurities  act as focal points of the spectral flow. 
\end{abstract}
\maketitle

\section{Introduction}

The stability of non-relativistic hydrogenoid atoms is one of the essential features that contributed to  consolidate  the  
quantum theory. However in  relativistic quantum mechanics there is a critical value of the central point-like 
charge $Ze$  from where on atom stability is lost \cite {Pomeran}-\cite{Popov1}. 
This is one of the surprising consequences of relativistic invariance in QED.  The phenomenon can
be understood  in a  heuristic way  as a falling to the center catastrophe. The critical value in  QED 
is reached when the spectrum of bound states of Dirac equation becomes complex which occurs for  $Z>137$. 
In fact what happens in QED is that when one of the bound states reaches the negative continuum spectrum the 
vacuum becomes unstable, generating electron-positron pairs. The positron escapes to infinite and the 
electron screens the central charge.
The phenomenon has attracted attention from a fundamental viewpoint  because it suggests that could be
experimentally tested by detecting an excess of positron
in the collisions of heavy nuclei \cite{Muller}\cite{Shabaev}. 

The instability of the atom for supercritical charges has also inspired a new mechanism of quark confinement in QCD 
\cite{Gribov2}--
\cite{Gribov5}.
The running of the strong coupling under  the renormalization group flow in QCD leads to large
values of the effective charge of the quarks, which reaches  very fast supercritical values in the infrared. 
The instability of the vacuum generates a transition 
from the perturbative Coulomb regime at short distances to a confinement regime in quarks interactions at large distances 
\cite{AS}--
\cite{AS5}.%

The discovery of graphene  \cite{Geim-Novoselov} opened a new window for the analysis of this 
phenomenon  \cite{Castro}. 
In that case a similar phenomenon occurs in the presence of charged impurities, but with a much lower  critical charge. In graphene the instability yields to a screening of the charge impurity, and  the phenomenon has been recently experimentally observed \cite{Wang0, Wang}. Motivated by this new physical effect  we review the
main features of this phenomenon and shed some light in some of its more paradoxical  aspects.
 We address the problem from a viewpoint where quantum unitarity is never lost no matter the strength of charge impurities. 
 In fact we show that  the formal analytic continauation of the bounded energy levels  of the Coulomb problem into  complex values does not mean a loss of Hermiticity of the corresponding effective Hamiltonian. It only shows the existence of
non-trivial spectral densities in the continuum spectrum which correspond to the existence of resonances in scattering
processes \cite{Wang0,Wang}. 

 In order to  clarify this issue we analyze   in graphene the transition from the subcritical regime   to the supercritical one by increasing the values of impurity charges. 
 The results show a continuous  behavior of the corresponding energy levels, although the spectral flow is very peculiar:  energy levels of Hydrogenoid spectrum in the critical regime are focal  points of the spectra
of subcritical and supercritical regimes. The peculiar behavior of the supercritical regime is reflected by the increasing  number of energy levels inside the energy gap, but the continuity of the spectral flow is always preserved along the transitions between  the different spectral regimes. Vacuum instability of the corresponding quantum field theory is pointed out   by the crossing of the  $E=-m$ energy level  of the Dirac sea continuum by some eingenvalues of the Dirac Hamiltonian which implies the appearance of pair particle-antiparticle creation mechanism that leads to the screening of the charge impurity.

The analysis of the problem is based in a  novel  method of dealing with selfadjoint extensions of the Dirac Hamiltonian. In that formalism all cases  are approached in an  unified and global way that allows to follow the spectral flow of the different (weak-strong) regimes in a smooth way. 
The analysis can be extended to any space dimension, e.g see  \cite{vor} for the three dimensional case.

In the  Section 2 we analyze the unitarity problem of the Dirac Hamiltonian in a Coulomb background.
The problem is solved by using the theory of self-adjoint extensions which regularize the singularities
associated to the Coulomb potential. The  selfadjoint Hamiltonians  are classified in different regimes according to the value of impurity charges. In Section 3 we calculate the bound states energy spectrum of
the Coulomb Hamiltonian in the different regimes. 
A particular attention is paid to the special cases of  Hydrogen and  {\it meta-Hydrogen} spectra (see \cite{hydrino}\cite{dom}), 
  The spectral flow of the bound states spectrum 
is  analysed in Section 4, where we also study the  analytic properties of this  flow
 in the different subcritical and critical regimes. Finally, the analysis of the results and conclusions is carried out in Section 5.

\section{Charged impurities in graphene}

Graphene is a two dimensional layer  of carbon atoms arranged on a honeycomb lattice of hexagons. 
The magic of the hexagonal honeycomb structure of graphene leads to a spectral structure in the
first Brillouin zone with two contact points  $K$ and $K'$ (Dirac points) between electronic  bands. In a neigbourghood 
of any of these two points \footnote{For simplicity we consider only the $K$ Dirac point  and $\hbar=1$}
 the  spectrum  of unbounded electrons  is well described   in terms of a massless Dirac Hamiltonian  \cite{Semenoff}
\begin{equation}
H=-i v_F (\sigma_1\partial_x+\sigma_2\partial_y),
\label{dirac}
\end{equation}
where $\sigma_i$, i=1,2,3, are the Pauli matrices and $v_F$ is the  velocity of the electrons at the Fermi surface, which for suspended graphene is about 300 times smaller than the  speed of light in vacuum.
This behavior also holds for graphene in a substrate of SiO$_2$ with a slight modification of $v_F$.

Although natural graphene behaves like a semi-metal with  no spectral gap, 
for  electronic applications it is convenient to open a gap between the bands
to reach a semiconductor regime. This behavior can be attained by different methods, either by introducing some disorder or by epitaxially grow graphene on a SiC substrate  \cite{Zhou}. In that case the effective Hamiltonian (\ref{Ham0}) becomes a massive Dirac Hamiltonian 
 \begin{equation}
H_0=-i v_F(\sigma_1\partial_1+\sigma_2\partial_2)+m\sigma_3,
\label{mdirac}
\end{equation}
where $m$ the effective mass of the gap.

In  the presence of a charged Coulomb impurity the effective electronic Hamiltonian 
becomes 
\begin{equation}
H_m=-i(\sigma_1\partial_x+\sigma_2\partial_y	)+m\sigma_3-\frac{\alpha}{r},
\label{Ham}
\end{equation}
where
$$
 \alpha=\frac{e_{*}^2}{v_F},\quad e_{*}^2=\frac{2e^2}{\epsilon+1},
$$
is the effective charge of the impurity,  
 $\epsilon$  the effective  dielectric constant of the graphene sheet,
 $r=\sqrt{x^2+y^2}$ and  the electronic effective speed factor $v_F$ has been absorbed by rescaling of
coordinates $x=x_1/v_F$, $y=x_2/v_F$. The values of $ \alpha$ depend on the substrate where the graphene sheet is grown. For instance,
$\alpha\approx 2$ for vacuum,  $ \alpha\simeq1$  for $\mbox{SiO}_2$ and $ \alpha\simeq0.35$ for $\mbox{SiC}$. 

The presence of a charge impurity with strong Coulomb interactions generate remarkable effects in the spectroscopic and transport properties. The physics of the effective theory is quite similar to that of relativistic atomic physics where the presence of instabilities is rather well know \cite {Pomeran}-\cite{Popov1}.
In any case there is a renewed interest  on the theoretical and experimental studies
on the Coulomb potential supercritical instabilities \cite{Novikov}-
\cite{Kotov}. The main difference with respect to the 3D analogue (Hydrogen-like atoms) is that the value of the supercritical charge is much smaller $\alpha=\frac12\ll 137$.

Although the single particle approach to the  Coulomb problem constitutes the first step in addressing nontrivial features of the full-fledged many-body interacting theory, most of the phenomenology of graphene physics can be explained from this simplified approach.

\subsection{2D Dirac Hamiltonian in a Coulomb background}

The  presence of a singularity at the origin of the Coulomb potential requires the use of some ultraviolet renormalization mechanism. 
 For such a reason  it is convenient to introduce an ultraviolet  cut-off $r_< >0$ around that singular point $r=0$ and later on take the appropriate limit  to extend the Hamiltonian to the whole space $\mathbb{R}^2\backslash \{0\}$ \cite{Pomeran,Case,Khalilov}. The only
 physical requirement is unitarity of time evolution, which is equivalent to the self-adjointness of the Hamiltonian defined in such limit.
If we exclude  from the physical space a  disk  $D(r_<)=\{x\in\R^2; \parallel x\parallel < r_<\}$ of  radius  $r_<$ around the origin, the most general boundary conditions can be given by \cite{ABPP,ABPP2,AIM2}
\begin{equation}
(1+\widehat{\rlap{n} /})\psi(r_<)=U(r_<)\sigma_3(1-\widehat{\rlap{n} /})\psi(r_<),
\label{b.c.}
\end{equation}
in terms of a unitary operator $U(r_<)$  defined on  the boundary values of spinors $\psi(r_<) \in L^2(S_{r_<}^1,\mathbb{C}^4)$,
where $\hat{n}$ denotes the normal vector to the circumference $S_{r_<}^1=\{x\in\R^2; \parallel x\parallel < r_<\}$.

Using polar coordinates $r$ and $\theta$ a general spinor $\psi$ can be expanded as
$$\psi(r,\phi)=\sum_{l=-\infty}^{\infty}(F_l(r)\Phi_l^+(\phi)+G_l(r)\Phi_l^-(\phi)),$$
in terms of  orthogonal eigenfunctions
$$
\Phi_l^{+}(\phi) =
\left( \begin{array}{c}
e^{i\,l\,\phi} \\
0
\end{array} \right) \mbox{ and }
\Phi_l^{-} (\phi)=
\left( \begin{array}{c}
0 \\
i\,e^{i\,(l+1)\,\phi}
\end{array} \right),
$$
of the total angular momentum $J_z=L_z+S_z=-i\frac{\partial}{\partial\phi}+\frac{1}{2}\sigma_3$, with semi-integer eigenvalues $j=l+1/2$. The space of spinors can be then decomposed as orthogonal sum of subspaces with fixed total angular momentum $j\in\Z+\half$:
\begin{equation}
\psi=\sum_j \psi_j ,
\label{decomposition}
\end{equation}
 where $\psi_j$ is a spinor of the form 
 \begin{equation}
\psi_j(r,\phi) =
\left( \begin{array}{c}
F^j(r)e^{i\,(j-1/2)\,\phi} \\
i G^j(r)e^{i\,(j+1/2)\,\phi}
\end{array} \right),
\label{psij}
\end{equation}
which  belongs to the subspace of total angular momentum $j$,

In order to preserve the $SO(2)$ rotation symmetry in the regularized theory, the unitary operator $U(r_<)$ fixing the boundary condition has to be diagonal in the angular momentum decomposition (\ref{decomposition}),
\begin{equation}
U(r_<)=\begin{pmatrix}
    \diagentry{\xddots}\\
    &\diagentry{e^{2i\,\beta_{j-1}}}\\ 
    &&\diagentry{e^{2i\,\beta_j}}\\
    &&&\diagentry{e^{2i\,\beta_{j+1}}}\\
    &&&&\diagentry{\xddots}\\
   \end{pmatrix}
\label{matrix}
\end{equation}
 i.e.  on each subspace of fixed angular momentum $j$ the unitary operator $U(r_<)$ reduces to a single phase $e^{2i\,\beta^j}$. Thus, the boundary condition (\ref{b.c.}) becomes
\begin{equation}
(1+\widehat{\rlap{n} /})\psi_j(r_<)=e^{2i\,\beta^j(r_<)}\sigma_3(1-\widehat{\rlap{n} /})\psi_j(r_<).
\label{b.c. subspace}
\end{equation}
More explicitly,
\begin{equation}
e^{2i\,\beta^j}=\frac{F^j(r_<)+i G^j(r_<)}{F^j(r_<)-i G^j(r_<)},
\label{beta0}
\end{equation}
where $F(r_<)$ and $G(r_<)$ are real functions.

The removal of  the UV regularization requires to take the limit $r_<\to 0$ which implies  the choice of
an appropriate series of boundary conditions $U(r_<)$.
The optimal choice of boundary conditions $U(r_<)$ that 
guarantees the convergence of the UV limit is given by the flow driven by {\it asymptotic zero modes}.
 Near the impurites asymptotic zero modes are solutions of the equation
\begin{equation}
\left[-i(\sigma_1\partial_x+\sigma_2\partial_y	)-\frac{ \alpha}{r}\right] \psi_0=0.
\label{Ham0}
\end{equation}
They will play a fundamental role in the renomalization of the singularity introduced by the impurities as they do in the
three-dimensional case of Hydrogenoid atoms \cite{vor}\cite{Gitman1}\cite{Book}.  The key observation is that 
in the  vicinity of the  inpurity $0<r\ll r_<$ any solutions of the Coulomb-Dirac equation $H\psi=E\psi$ behaves as an asymptotic zero mode.  Thus,  all the spinors in the domain of the Hamiltonian must behave  near the singularity as zero modes of (\ref{Ham0}).

 For any  choice of boundary condition $\beta^j_0$  at a given cut-off $r_0\ll1$ 
 there is a unique asymptotic zero mode $(F_0^j, G_0^j)$ satisfying the equation
  \begin{equation}
e^{2i\,\beta_0^j}=\frac{F_0^j(r_0)+i G_0^j(r_0)}{F_0^j(r_0)-i G_0^j(r_0)}.
\label{zeromode}
\end{equation} 
If the two components of the asymptotic zero mode $(F_0^j, G_0^j)$  are $L^2$ normalizable   in a neigbourghood of the singularity, i.e. $F^j_{0},G^j_{0} \in L^2(D(r_<),\mathbb{C})$, then the flow  of boundary conditions  $\beta^j_{r_<}$ 
($r_<\in (0, r_0)$) given by  
\begin{equation}
 e^{2i\,\beta^j_{r_<}}=  \frac{F^j_{0}(r_<)+i G^j_{0}(r_<)}{F^j_{0}(r_<)-i G^j_{0}(r_<)}.
 \label{zeromode+}
\end{equation} 
 defines in the limit $r_<\to 0$ 
a selfadjoint extension of  the Dirac Hamiltonian \eref{Ham}. The  domain of the Hamiltonian is  expanded by the
 spinors $(F^j_{0},G^j_{0}) $ which satisfy
\begin{equation}
\lim_{r\rightarrow 0}\Big(F^j(r)G^j_{0}(r)-G^j(r)F^j_{0}(r)\Big)=0.
\label{bound cond}
\end{equation}

In other terms, once the cut-off $r_0$ is fixed,  we can associate to each boundary condition parametrized by $\beta_0^j$ a unique asymptotic zero mode satisfying (\ref{zeromode}). On the other way, given an asymptotic zero mode, the relation (\ref{zeromode+}) defines for each  $r_<\in (0,r_0)$ a boundary phase $\beta^j_{r_<}$ in a unique way.  

For some values of the impurity charge not all the boundary conditions  $\beta_0^j$ give rise to  normalizable asymptotic zero modes. Such boundary conditions do not lead by the procedure described above to a  well defined selfadjoint Dirac Hamiltonian. However, as we shall see later on, it is always possible to find an alternative boundary condition $\widehat{\beta}_0^{j}$ for the same value of impurity charge  whose zero mode is normalizable and leads to a well defined selfadjoint Dirac Hamiltonian \footnote{In any case one can define alternative prescriptions of  the boundary conditions flows which starting from a non-normalizable  boundary condition converge to the trajectories of normalizable boundary conditions. However, these prescriptions are not canonical and  will  not be consider here}.

By this  method we have replaced the convergent flow of   UV cut-off boundary conditions just by the choice of a simple 
asymptotic boundary condition  \eref{bound cond}. The boundary condition flow is then defined in this way:  the initial cut-off phase $\beta_0^j$ defines an asymptotic zero mode $(F^j_{0},G^j_{0})$, and the boundary phases $\beta^j(r_<)$ run with the cut-off while keeping fixed the zero mode, converging to a well defined boundary condition when the cut-off is removed.

In summary, the boundary condition of the Dirac Hamiltonian in a  Coulomb background is defined by the choice of one of these  two equivalent boundary data: either a 
unitary matrix $U(r_0)$ of the form \eref{matrix} acting on the functions of  the boundary of a small cut-off disk of radius $r_0\ll1$  or a normalizable  asymptotic zero mode $(F^j_{0},G^j_{0})$. The connection between the two choices is given by equation \eref{b.c.}.
Moreover, any boundary condition that leads to selfadjoint extension of the Hamiltonian  \eref{Ham}  is obtained by this method.

\subsection{Boundary conditions for different regimes}
 The subspace of asymptotic zero modes $(F^j_{0},G^j_{0})$ satisfying the boundary condition  \eref{beta0} for a given angular momentum $j\in \Z +\half$ 
depends on the value of the charge $ \alpha$ of the impurity, in a similar way as in the three-dimensional analogue case \cite{vor,Gitman1,Book}

To find the asymptotic zero modes of the Hamiltonian (\ref{Ham}) we have to look only at  leading terms
asymptotic expansion around the impurity.
Using the expansion  (\ref{psij}) is  easy to show  that they 
satisfy the following coupled equations
\begin{equation}
\frac{d F^j_{0}}{dr}-\frac{j-1/2}{r}F^j_{0}+\frac{ \alpha}{r}G^j_{0}=0,
 \label{eq F 0}
\end{equation}
\begin{equation}
\frac{d G^j_{0}}{dr}+\frac{j+1/2}{r}G^j_{0}-\frac{ \alpha}{r}F^j_{0}=0.
 \label{eq G 0}
 \end{equation}
Searching for solutions of the form $F^j_{0}(r)=r^s$ and $G^j_{0}(r)=C\,r^s$ we find two independent solutions
\begin{equation}
s_\pm=-1/2\pm\nu, \quad C_\pm=(j\mp\nu)/ \alpha \qquad \hbox{if}\quad  \alpha^2\neq j^2,
\end{equation}
where $\nu=\sqrt{j^2-  \alpha^2}$. 
For $ \alpha^2= j^2$ the two solutions degenerate, but in this case the logarithmic corrections give rise  also  to two independent 
solutions of the form%
\begin{eqnarray}
F^j_{0}(r)= &r^{-1/2}\qquad\quad &G^j_{0}(r)=\frac{j}{|j|}\, r^{-1/2} \\
F^j_{0}(r)= &r^{-1/2}\log(\Lambda\,r) & G^j_{0}(r)=\frac{j}{|j|}\, r^{-1/2}\left[\log(\Lambda\,r)-\frac1{j}\right],
\end{eqnarray}
where $\Lambda$ is a parameter with dimensions of momentum $[L]^{-1}$.

Notice that the value $j^2= \alpha^2$ is critical: when $\alpha^2<j^2$ the parameter $\nu$ is real, while for $\alpha^2>j^2$ it is  purely imaginary.  Thus, depending on the strength of the charge impurity there are three different regimes where to impose the boundary conditions. 

\bigskip
{\bf\large a) Regular regime: $\alpha^2\leq j^2-\frac14$}
\bigskip

This regime is never reached in the lowest angular momentum states $j=\pm1/2$.

In this case $\nu$ is  a real parameter and one of the two asymptotic zero modes solutions is not normalizable
in a neigbourghood of the origen $D(r_0)$. Indeed,  the asymptotic zero mode
\begin{equation}
\psi^-_j(r,\phi) =
\left( \begin{array}{c}
r^{s_-}\, e^{i\,(j-1/2)\,\phi} \\
i C_-r^{s_-}\, e^{i\,(j+1/2)\,\phi} 
\end{array} \right)  \notin L^2(D(r_0),\mathbb{C}^2).
\label{psij-}
\end{equation}
is not square integrable in $D(r_0)$.
Thus, we are left with only one asymptotic behaviour given by the normalizable zero mode
\begin{equation}
\psi^+_j(r,\phi) =
\left( \begin{array}{c}
r^{s_+}\, e^{i\,(j-1/2)\,\phi} \\
i C_+r^{s_+}\, e^{i\,(j+1/2)\,\phi} 
\end{array} \right)  \in L^2(D(r_0),\mathbb{C}^2),
\label{psij+}
\end{equation}
which strongly constrains  the boundary condition (\ref{beta0}), 
In particular, the parameter $\beta^j_0$ is completely fixed, independently from $r_0$, by
$$
\beta^j_0=\half\arcsin\displaystyle \left| \frac{\alpha}{j}\right|.
$$
This means that there is a unique self adjoint extension of the Hamiltonian \eref{Ham}. 
The boundary condition (\ref{bound cond}) becomes: 
\begin{equation}
\lim_{r\rightarrow 0}\Big[(-j+\nu)F^j(r)+ \alpha G^j(r)\Big]=0.
 \label{bc i}
 \end{equation}

\bigskip
{\bf\large b) Subcritical regime: $j^2-\frac14<\alpha^2< j^2$}
\bigskip

In this regime both  solutions are normalizable, thus the most general asymptotic zero mode is a linear combination of the two solutions \eref{psij-} \eref{psij+}.  The choice of  $\beta^j_0\in[0,\pi)$  fixes that linear  combination in a unique way, up to  a global  constant. 
\begin{eqnarray}
F^j_0(r) &=&r^{-1/2}\big(\cos\theta\,(\Lambda\,r)^{\nu}-\sin\theta\,(\Lambda\,r)^{-\nu}\big)\nonumber\\
G^j_0(r) &=&r^{-1/2}\big(\cos\theta\,C_+(\Lambda\,r)^{\nu}-\sin\theta\,C_-(\Lambda\,r)^{-\nu}\big),
\label{F0 G0 case 1}
\end{eqnarray}
where the parameter $\theta\in[0,\pi)$  of the linear combination is given by
$$
\tan\theta=\frac{(j-\nu)\cos\beta^j_0- \alpha\sin\beta^j_0}{(j+\nu)\cos\beta^j_0- \alpha\sin\beta^j_0}(\Lambda\,r_0)^{2\nu}.
$$
Thus,  the boundary condition (\ref{bound cond}) becomes:
\begin{equation}
\lim_{r\rightarrow 0}\Big[\Big(j(\tan\theta-(\Lambda\,r)^{2\nu})+\nu(\tan\theta+(\Lambda\,r)^{2\nu})\Big) F^j(r)- \alpha(\tan\theta-(\Lambda\,r)^{2\nu}) G^j(r)\Big]=0
 \label{bc ii}
 \end{equation}
 
 \bigskip
{\bf\large c) Critical regime: $\alpha^2=j^2$}
\bigskip

In this case the most general asymptotic zero mode is 
\begin{eqnarray}
F^j_0(r) &=&r^{-1/2}\big(\cos\theta+\sin\theta\,\log(\Lambda\,r)\big)\nonumber\\
G^j_0(r) &=&r^{-1/2}\frac{j}{|j|}\Big(\cos\theta+\sin\theta\,\big(\log(\Lambda\,r)-\textstyle\frac{1}{j}\big)\Big),
\label{F0 G0 case 2}
\end{eqnarray}
where the parameter $\theta\in[0,\pi)$ can be related  to the phase  $\beta^j_0\in[0,\pi)$ of the boundary condition (\ref{beta0})   imposed at $S^1_{r_0}$
$$
\tan\theta=\frac{j-|j|\tan\beta^j_0}{1-\log(\Lambda\,r)(j-|j|\tan\beta^j_0)}.
$$
The corresponding boundary condition (\ref{bound cond}) of the Dirac operator  becomes:
\begin{equation}
\lim_{r\rightarrow 0}\Big[|j|\,G^j(r)\Big(1+\log (\Lambda r)\tan\theta\Big)-F^j(r)\Big(j+\big[-1+j\,\log (\Lambda r)\big]\tan\theta\Big)\Big]=0.
 \label{bc iii}
 \end{equation}

  \bigskip
{\bf\large d) Supercritical regime: $\alpha^2>j^2$}
\bigskip

In this case the value of $\nu$ becomes imaginary and 
both asymptotic zero modes are normalizable. A  general zero mode solution is of the form
\begin{eqnarray}
F^j_0(r) &=&r^{-1/2}\big(e^{-i\,\theta}\,(\Lambda\,r)^{\nu}+ e^{i\,\theta}\,(\Lambda\,r)^{-\nu}\big)\nonumber\\
G^j_0(r) &=&r^{-1/2}\big(e^{-i\,\theta}\,C_+(\Lambda\,r)^{\nu}+ e^{i\,\theta}\,C_-(\Lambda\,r)^{-\nu}\big),
\label{F0 G0 case 3}
\end{eqnarray}
where the parameter $\theta\in[0,\pi)$ is fixed by the phase  $\beta^j_0\in[0,\pi)$ of the boundary condition (\ref{beta0})  imposed at $S^1_{r_0}$
$$
 e^{2\theta i }=\frac{(\nu-j)\cos\textstyle\beta^j_0+ \alpha\sin\beta^j_0}{(\nu+j)\cos\beta^j_0- \alpha\sin\beta^j_0}(\Lambda r_0)^{2\nu}.
$$
The asymptotic boundary condition (\ref{bound cond}) in this case reads
\begin{equation}
\lim_{r\rightarrow 0}\Big[\Big(\nu( e^{2i\theta}-(\Lambda\,r)^{2\nu})+j( e^{2i\theta}+(\Lambda\,r)^{2\nu})\Big) F^j(r)- \alpha \ (e^{2i\theta}+(\Lambda\,r)^{2\nu}) G^j(r)\Big]=0.
 \label{bc iv}
 \end{equation}
 Notice that in any of the above regimes  the Hamiltonian \eref{Ham} is a selfadjoint operator.
 
From now on we will parametrize the boundary conditions by $\theta\in(0,\pi)$ 
keeping in mind its relations with $\beta^j_0$ and $r_0$.

\section{Bound states and energy levels}
Once we have shown that the Dirac Hamiltonian  \eref{Ham} is a selfadjoint operator it is possible
to analyze its spectrum by   finding the energy levels
\begin{equation}
H\psi=E\psi.
\label{eig pro}
\end{equation}
The eigenvalue problem can be reduced, by using the ansatz (\ref{psij}) for each subspace of fixed angular momentum $j$,  to solve the pair of  coupled differential equations
\begin{equation}
\frac{dF^j}{dr}-\frac{j-1/2}{r}F^j+(E+m+\frac{ \alpha}{r})G^j=0,
 \label{eq F}
\end{equation}
\begin{equation}
\frac{dG^j}{dr}+\frac{j+1/2}{r}G^j-(E-m+\frac{ \alpha}{r})F^j=0.
 \label{eq G}
\end{equation}

Let us now  introduce two radial functions $a(r)$ and $b(r)$
defined by
\begin{equation}
F^j(r)=\frac{\sqrt{m+E}}{2\,r}\big(a(r)-b(r)\big),\,\,G^j(r)=\frac{\sqrt{m-E}}{2\,r}\big(a(r)+b(r)\big),\,\,,
 \label{replacements}
\end{equation}
and use the notation $\epsilon=\sqrt{m^2-E^2},\,\,x=2\epsilon\,r$.
 The coupled equations satisfied by  $a(x)$ and $b(x)$ become
\begin{equation}
xa'(x)+\Big(\frac{x}{2}-\frac{1}{2}-\frac{ \alpha E}{\epsilon}\Big)a(x)+\Big(\frac{ \alpha m}{\epsilon}+j\Big)b(x)=0, \label{ab1}
\end{equation}
\begin{equation}
xb'(x)-\Big(\frac{x}{2}+\frac{1}{2}-\frac{ \alpha E}{\epsilon}\Big)b(x)+\Big(-\frac{ \alpha m}{\epsilon}+j\Big)a(x)=0,\label{ab2}
\end{equation}
which can be easily decoupled. Indeed it is obvious to realize that  
\begin{equation}
b(x)=\frac{(2 \alpha E+\epsilon-\epsilon\,x)a(x) -2\epsilon\,x\,a'(x))}{2( \alpha m+j \epsilon)}
\label{b}
\end{equation}
and then
\begin{equation}
a''(x)+\Big(\textstyle -\frac{1}{4}+\frac{\frac{1}{2}+ \alpha\frac{E}{\epsilon}}{x}+\frac{\frac{1}{4}-j^2+\alpha^2}{x^2}\Big)a(x)=0,\label{a}
\end{equation}
The general solution of \eref{a} can be expressed in terms of Whittaker functions$W$and M \cite{Abramo}
\begin{equation}
a(x)=A\, W(1/2+ \alpha E/\epsilon,\nu,x) + B\, M(1/2+ \alpha E/\epsilon,\nu,x),
\label{sol a}
\end{equation}
where A and B are constants. In the same way we have that
\begin{equation}
b(x)=(j- \alpha m/\epsilon)\,A\,W(-1/2+ \alpha E/\epsilon,\nu,x) + \Big(\frac{ \alpha m -j\epsilon}{ \alpha E +\nu \epsilon}\Big)\,B\, M(-1/2+ \alpha E/\epsilon,\nu,x).
\label{sol b}
\end{equation}
Thus,  the  general solution of (\ref{eq F}) and (\ref{eq G}) is given by
\begin{eqnarray}
F^j(r)&=&\textstyle \frac{\sqrt{m+E}}{2\,r}\Bigg[A\Big(W(1/2+ \alpha E/\epsilon,\nu,2\epsilon r)-(j- \alpha m/\epsilon)W(-1/2+ \alpha E/\epsilon,\nu,2\epsilon r)\Big) +\nonumber\\
& &\textstyle +B\Big(M(1/2+ \alpha E/\epsilon,\nu,2\epsilon r)-\Big(\frac{ \alpha m -j\epsilon}{ \alpha E +\nu \epsilon}\Big)M(-1/2+ \alpha E/\epsilon,\nu,2\epsilon r)\Big)\Bigg], 
\\
G^j(r)&=&\textstyle\frac{\sqrt{m-E}}{2\,r}\Bigg[A\Big(W(1/2+ \alpha E/\epsilon,\nu,2\epsilon r)+(j- \alpha m/\epsilon)W(-1/2+ \alpha E/\epsilon,\nu,2\epsilon r)\Big) +\nonumber\\
& &\textstyle+B\Big(M(1/2+ \alpha E/\epsilon,\nu,2\epsilon r)+\Big(\frac{ \alpha m -j\epsilon}{ \alpha E +\nu \epsilon}\Big)M(-1/2+ \alpha E/\epsilon,\nu,2\epsilon r)\Big)\Bigg].
\label{sol F G}
\end{eqnarray}

The  asymptotic behavior of these solutions is strongly dependent on the regime of charge impurities.

 If  $\alpha^2\neq j^2$  the asymptotic behavior can be derived from the behavior of the   
Whittaker functions $W$ and $M$ for $r\ll1$, 
\begin{eqnarray}
M(\pm\half+ \alpha E/\epsilon,\nu,x)&\cong&x^{1/2+\nu}\\
W(\pm\half+ \alpha E/\epsilon,\nu,x)&\cong&x^{1/2}\Big(\frac{x^{\nu}\,\Gamma[2\,\nu]}{\Gamma[\half\mp\half-\nu- \alpha E/\epsilon]}+\frac{x^{-\nu}\,\Gamma[2\,\nu]}{\Gamma[\half\mp\half+\nu- \alpha E/\epsilon]}\Big),
\label{exp W}
\end{eqnarray}
whereas for $\alpha^2=j^2$:
\begin{eqnarray}
M(\pm1/2+ \alpha E/\epsilon,0,x)&\cong&x^{1/2}\\
W(\pm1/2+ \alpha E/\epsilon,0,x)&\cong&-\frac{x^{1/2}}{\Gamma[\half\mp\half-j E/\epsilon]}\Big(2\gamma+\psi(\half\mp\half-j E/\epsilon)+\log x\Big),
\label{exp W aj}
\end{eqnarray}
where 
$\psi(x)=\Gamma'(x)/\Gamma(x)$ is the digamma  function and $\gamma$ the Euler's constant.

The spectrum of energy levels is  also strongly  dependent on  the regime of charges. Let us 
focus on the discrete energy spectrum which correspond to bound states.

\subsection{Regular regime $\alpha^2\leq j^2-\frac14$} \label{regular}

The boundary conditions (\ref{bc i}) can only  be satisfied if the constant $A$ of the general solution (\ref{sol F G})  vanishes. On the other hand bound state spinors $\psi$ have to be $L^2(\R^3,\mathbb{C}^2)$-normalizable which implies that it must to  decay at infinity. Thus, the asymptotic behaviour   at $r\gg1$ of (\ref{eq F}) and (\ref{eq G})  must be of the form $\big(F^j(r)\cong\, e^{-\epsilon r}$,$\,G^j(r)\cong\, e^{-\epsilon r}\big)$. The implies that the spinors should look like $$F^j(r)=r^{-1/2+\nu}\, e^{-\epsilon r}f(r),\,\,\, G^j(r)=r^{-1/2+\nu}\, e^{-\epsilon r}g(r),$$ where  $f(r)$ and $g(r)$ are two radial functions that are polynomially bounded at infinity.  This requirement is satisfied when the expressions $$P_1(r)=r^{-1/2-\nu}e^{\epsilon r}M(1/2+ \alpha E/\epsilon,\nu,2\epsilon r),$$ 
$$P_2(r)=r^{-1/2-\nu}e^{\epsilon r}M(-1/2+ \alpha E/\epsilon,\nu,2\epsilon r)$$ reduce  to polynomials, or 
when only $P_1(x)$  is a polynomial and  $ \alpha m =j\epsilon$. Expanding $P_1$ and $P_2$ it is possible to show that this happens when $- \alpha E/\epsilon+\nu=-n$, with $n=0,1,2,..$ if $j>0$ and $n=1,2,..$ if $j<0$. More explicitly, the
spectrum of bound states is given by \\
\begin{eqnarray}
E^H_n=\frac{m}{\sqrt{1+\frac{\alpha^2}{(n+\sqrt{j^2-\alpha^2})^2}}}, & \,\,\, &
n=\left\{ \begin{array}{l}
0,1,2,..\mbox{ for } j>0 \\
1,2,3,..\mbox{ for } j<0
\end{array}. \right.
\label{eig 1}
\end{eqnarray}
This is the well known Hydrogenoid atom spectrum of  bound states. 

For $\alpha^2>j^2-\frac14$ the boundary conditions (\ref{bc ii}) and (\ref{bc iii}) for $\theta\neq0$ and $\theta\neq\frac{\pi}{2}$ (we will analyze these two exceptional cases later separately), and (\ref{bc iv}) (for any value of  $\theta$) are satisfied only if  the parameter $B$ of the general solution (\ref{sol F G}) vanishes $B=0$. In that case only terms involving the Whittaker function $W$ survive, which implies that they automatically decays exponentially at infinity.

In this sense  the exponential decay $e^{-\epsilon r}$ of bound states means that they are localized around the impurity charge and thus behave as edge states in topological insulators \cite{ABPP,ABPP2}.

Notice that in the massless limit the exponential decay $e^{-\epsilon r}$ becomes a pure phase factor $e^{-i E r}$ and the corresponding solution is not localized and in fact belongs to the continuum energy spectrum.

\subsection{Subcritical regime  $j^2-\frac14<\alpha^2<j^2$ with $\theta\neq0$  and $\theta\neq\frac{\pi}{2}$} \label{subcritical}
In that case   the boundary conditions (\ref{bc ii})  are satisfied  only if  
\begin{equation}
\frac{\big( \alpha(-E+m)+(-j+\nu)\epsilon\big)}{\big( \alpha(E-m)+(j+\nu)\epsilon\big)}\frac{\Gamma[2\,\nu]\Gamma[1-\nu- \alpha E/\epsilon]}{\Gamma[-2\,\nu]\Gamma[1+\nu- \alpha E/\epsilon]}=\Big(\frac{\Lambda}{2\,\epsilon}\Big)^{-2\,\nu}\tan\theta.\label{eig 3a}
\end{equation}
The solution of \eref{eig 3a} gives the spectrum  $E_n^{II}(\theta)$ of bound states in the
subcritical regime.
\subsection{Critical regime  $\alpha^2=j^2$, for  $\theta\neq0$  and $\theta\neq{\pi}$}  \label{critical}

In this case the spectral condition derived from the boundary conditions  (\ref{bc iii})  is
\begin{equation}
\frac{j-|j|\,{(m_{\epsilon}-E_{\epsilon})}}{(j-{|j|\,(m_\epsilon-E_{\epsilon})})(2\gamma -\log\Lambda_{\epsilon}/2)+ (j-{|j|\,m_{\epsilon}})\psi(1-{|j|\, E_{\epsilon}})+{|j|\,E_{\epsilon}}\psi(-{|j|\, E_{\epsilon}})}
=\tan\theta.
\label{eig 3b}
\end{equation}
where $E_{\epsilon}=E/\epsilon$, $m_{\epsilon}=m/\epsilon$ and $\Lambda_{\epsilon}=\Lambda/\epsilon$.
The solutions of equation \eref{eig 3b} give an  infinite sequence   $E_n^{III}(\theta)$ of discrete energy levels.

\subsection{Supercritical regime  $\alpha^2>j^2$}
In this case the infinite set of energy levels  $E_n^{IV}(\theta)$  $n\in \mathbb{Z}$ is given by the spectral condition
\begin{equation}
\frac{\big( \alpha(-E+m)+(-j+\nu)\epsilon\big)}{\big( \alpha(E-m)+(j+\nu)\epsilon\big)}\frac{\Gamma[2\,\nu]\Gamma[1-\nu- \alpha E/\epsilon]}{\Gamma[-2\,\nu]\Gamma[1+\nu- \alpha E/\epsilon]}=-e^{2\,i\,\theta}\Big(\frac{\Lambda}{2\,\epsilon}\Big)^{-2\,\nu}.\label{eig 3c}
\end{equation}
In the subcritical regime $j^2-\frac14<\alpha^2<j^2$ (\ref{subcritical}), we have two special cases:  $\theta=0$ and $\theta=\frac\pi2$ where the asymptotic 
zero modes that defines the boundary conditions are reduced to one of the two different asymptotic behaviors near the origin. 
\subsection{Subcritical regime  $j^2-\frac14<\alpha^2<j^2$ with $\theta=0$ (Hydrogenoid atom)} 
The boundary conditions reduce in this case those of the regular regime  (\ref{regular}) , 
and then the spectrum is the same as  in (\ref{eig 1}), i.e the bound states spectrum is the same as the
Hydrogenoid atom $E_n^H$.\\ \\
\subsection{Subcritical regime  $j^2-\frac14<\alpha^2<j^2$ with $\theta=\frac\pi2$ (Meta-Hydrogenoid atom)} 
If $\theta=\frac\pi2$   the boundary conditions are defined by the  asymptotic zero modes characterized by the exponent $s=-\frac12-\nu$ and become
\begin{equation}
\lim_{r\rightarrow 0}\Big(-(j+\nu)F^j(r)+ \alpha G^j(r)\Big)=0.
 \label{bc meta-Hydrogenoid}
 \end{equation}
These boundary conditions can be satisfied by setting $A=0$ in the general solutions and making the replacement $\nu\rightarrow-\nu$. Using the same techniques  as in the regular case, we get the analytic spectrum of bound states
\begin{equation}
\left\{ \begin{array}{l}
E^h_0=-\frac{m}{\sqrt{1+\frac{\alpha^2}{{j^2-\alpha^2}}}}\\
E^h_n=\frac{m}{\sqrt{1+\frac{\alpha^2}{(n-\sqrt{(j^2-\alpha^2})^2}}}
\end{array} \right.\\
\left. \begin{array}{l}
\mbox{ for } j>0,
\end{array} \right.\label{eig2p}
\end{equation}
and
\begin{equation}
\left. \begin{array}{l}
E^h_n=\frac{m}{\sqrt{1+\frac{\alpha^2}{(n-\sqrt{(j^2-\alpha^2})^2}}}
\end{array} \right.\\
\left. \begin{array}{l}
\mbox{ for } j<0,
\end{array} \right.
\label{eig 2}
\end{equation}
with $n=1,2,3,...$. The above bound states are known {\it meta-Hydrogenoid} states \footnote{The meta-Hydrogenoid states first appeared in the
 literature as hydrino states \cite{hydrino}\cite{dom}. However, the misuse of its properties for claiming magic generation of energy requires the
 introduction of new  name. Notice that the Hydrogen atom $Z=1, D=3$ is  in a subcritical regime where the 
 Hamiltionian is essentially selfadjoint and there is a canonical boundary condition giving rise to the well know spectrum.
 There is no meta-Hydrogen spectrum. Otherwise it will open the interesting window  to explain the puzzle of proton radius in an elegant way in terms of more exotic
 boundary conditions.
  }. 
  The   meta-Hydrogenoid bounded spectrum is very similar to the Hydrogenoid spectrum. The only difference is a sign in the second radical.

As in the Hydrogen case for $\alpha^2>j^2$ the spectral formulae  \eref{eig2p} \eref{eig 2} becomes complex, but as we have already remarked the real spectrum is given by \eref{eig 3c}.
\subsection{Critical regime  $\alpha^2=j^2$ with  $\theta=0$ (or $\theta=\pi$)}

In the critical regime for boundary conditions with $\theta=0$  the Hydrogenoid  and meta-Hydro--genoid spectra do coincide.
They are defined by (\ref{eig 1}) with the only difference that $E_0=0$ for $\alpha^2=j^2$ and $j>0$. 

The Hydrogenoid  and meta-Hydrogenoid spectra are not defined for $\alpha^2>j^2$.

\vspace{-.1cm}
\section{Spectral flows of bound states}

The problem which inspired the Gribov approach to confinement is the fact that  the energies of the bound states 
given by $E^H_n$  become complex for $\alpha^2>j^2$.  
To  better understand that mechanism let us analyse the flow
of the bound state spectrum by continuoulsy increasing the charge $\alpha$ of the  impurity 
or varying the boundary
conditions.

\subsection{Spectral flow and boundary conditions}

It is interesting to analyze the flow of the spectrum as we change the boundary condition parameter $\theta$.
The continuous flow of  the spectrum defined by the  change of the parameter $\theta$ characterizing the boundary conditions in the  {\it subcritical regime}  $j^2-1/4<\alpha^2< j^2$ is displayed  in Figure 1. There we plot  for $j=\frac32$ the $\theta$ dependence of the lowest energy bound states in this regime.

Notice that in the limits $\theta=0,\theta=\pi/2,\theta=\pi$ we recover the Hydrogenoid and meta-Hydrogenoid spectrum, i.e.
\begin{eqnarray}
\lim_{\theta\to0} E^{II}_n(\theta)&=&E^H_n,\label{S3S1}\\
\lim_{\theta\to\frac\pi2} E^{II}_n(\theta)&=&E^h_n\\
\lim_{\theta\to\pi} E^{II}_n(\theta)&=&E^H_{n+1}.
\end{eqnarray}

\begin{figure}[h!]
\centering{\includegraphics[width=14cm]{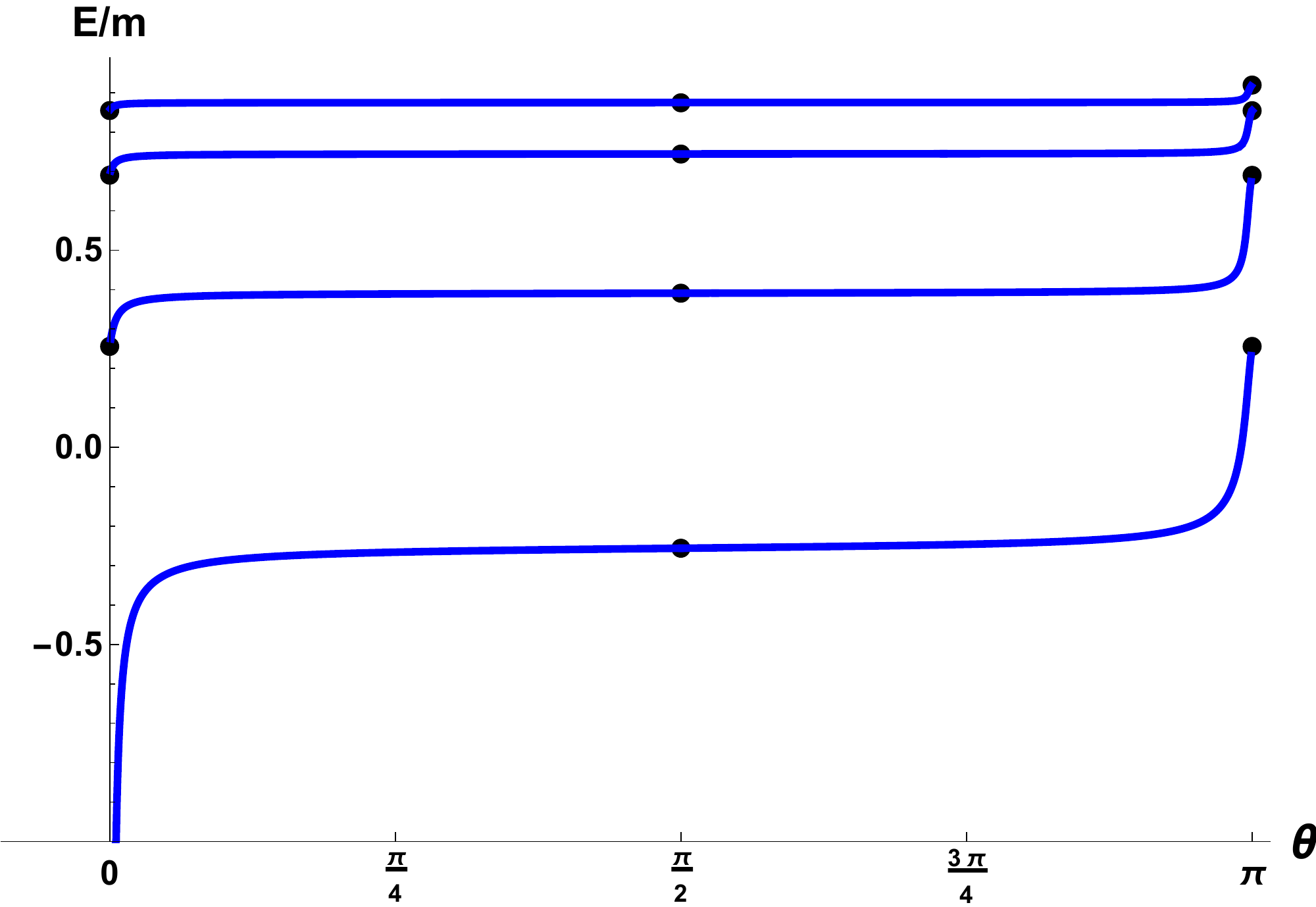}}
\caption{$\theta$ dependence of the energy ${E}/{m}$ of the lowest bound states with angular momentum $j=3/2$
for $ \alpha=1.45$ in the {\it subcritical regime} $2<\alpha^2< \frac94$ ($\Lambda=m/10$). 
 The dots correspond to the Hydrogenoid and meta-Hydrogenoid energy levels at $\theta=0$, $\theta=\frac\pi2$ and $\theta=\pi$.}
\label{fig sub}
\end{figure}
The continuity of the flow should be  obvious  from the fact that the boundary conditions (\ref{bc ii}) reduce to the boundary conditions of the Hydrogenoid and meta-Hydrogenoid spectra in these limits. What is more surprising is the fact that the spectral
flow is not periodic, i.e. there is an {\it spectral asymmetry}. The spectrum is periodic, i.e. it is the same at $\theta$ and $\theta+\pi$, but the flow shifts the energy levels by one unit in each cycle from $\theta=0$ to $\theta=\pi$. 
In general, for fixed angular momentum and charge
we have
\begin{equation}
 E^{II}_n(\theta+k \pi)=E^{II}_{n+k} (\theta),\label{period}
\end{equation}
for any integer $k\in \Z$. The behaviour of the spectral flow recalls the pumping mechanism exhibited by edge states in topological 
insulators \cite{Laughlin, Halperin}.

Another interesting property of the spectral flow  is its  monotoncity, i.e. $ E^{II}(\theta)< E^{II}(\theta')$ if 
$ \theta< \theta' $.
In particular this implies the standard sandwich inequalities between the Hydrogenoid and meta-Hydrogenoid energy levels
\begin{equation}
-m<E^h_{0}<E_{0}^{H}<E_{1}^{h}<E_{1}^{H}<\dots<E_{n-1}^{h}<E_{n-1}^{H}<E_{n}^{h}<E_{n}^{H}<\dots< m.\label{rel eig}
\end{equation}
Notice that there is a bound state emerging from the continuum $E\leq -m$ at a value of $\theta$ close to $\theta=0$.
The  behaviour of the flow is the same  for  positive $j>0$ and negative angular momentum  $j<0$, except for the
 absence of zero levels (n=0) for the Hydrogenoid and meta-Hydrogenoid energy levels for $j<0$ . Thus the   sandwich inequalities in the negative case are
 \begin{equation}
-m<E_{1}^{h}<E_{1}^{H}<\dots<E_{n-1}^{h}<E_{n-1}^{H}<E_{n}^{h}<E_{n}^{H}<\dots< m.\label{rel eig2}
\end{equation}
\begin{figure}[h!]
\centering{\includegraphics[width=14cm]{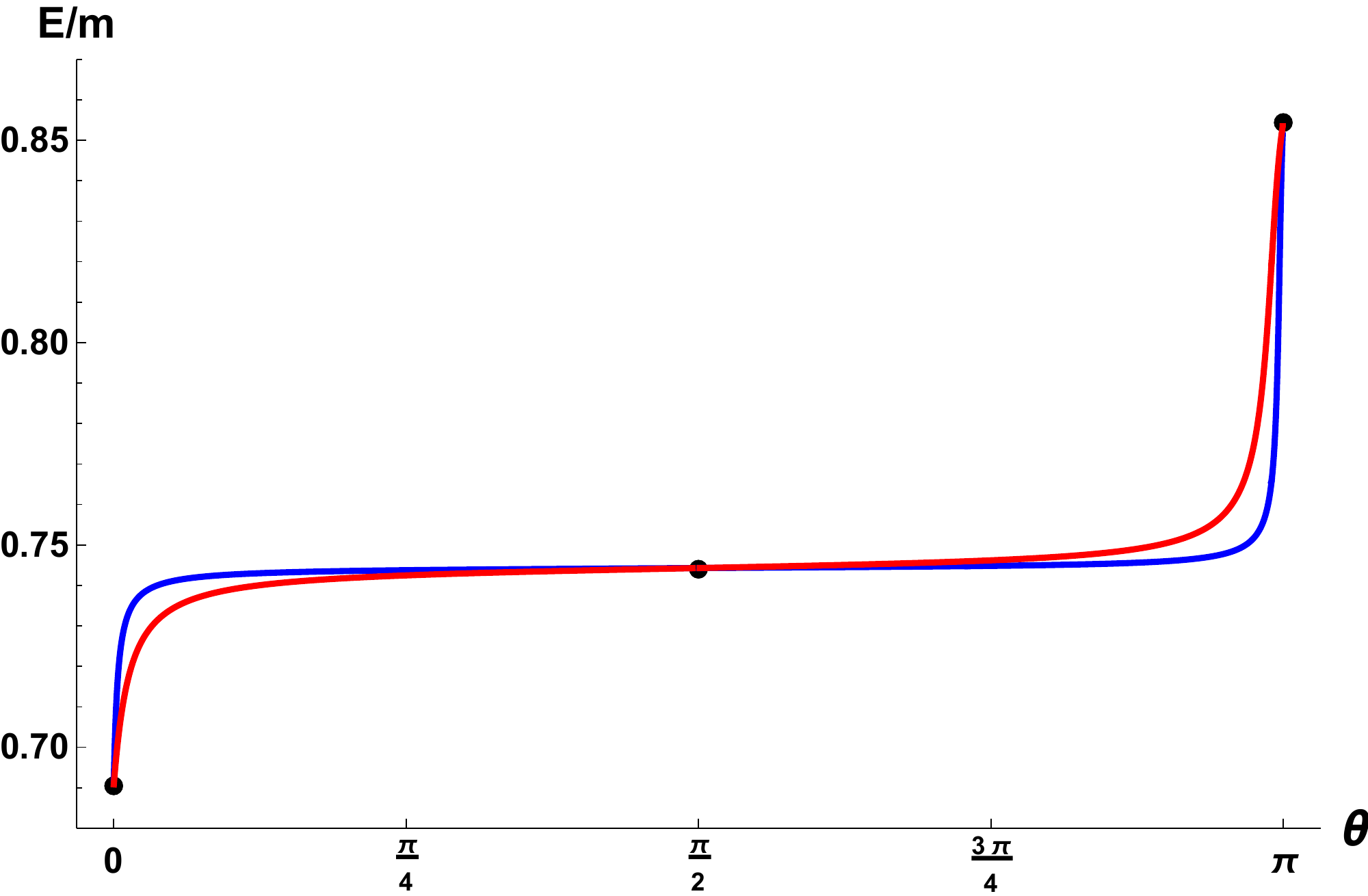}}
\caption{Gap between the energies corresponding to $j=3/2$ (blue) and $j=-3/2$ (red) for $\alpha=1.45$ and 
$\Lambda=m/10$ in the subcritical regime.}
\label{fig gap}
\end{figure}
 
Another interesting feature of  the subcritical regime is that  from (\ref{eig 1}) and (\ref{eig 2}) it follows  that for $n>0$ the spectra $E^H_n$ and $E^h_n$ with $j>0$ and $j<0$ are degenerate. The boundary condition (\ref{bc ii}) for $\theta\neq k \pi$ and $\theta\neq \frac{2k+1}2$ breaks this degeneracy and creates a gap between the energies corresponding to $j>0$ and $j<0$. The situation is described in  Figure \ref{fig gap} for $Z\alpha=1.45$. For $\theta=0$ we have the energy corresponding to $n=1$ of the Hydrogen spectrum $E^H_1$ which is degenerate for $j=\pm\frac32$. As we increase the parameter $\theta$ a gap appears between the states $j=\pm\frac32$. The energy of the state $j=-\frac32$ becomes lower than the one corresponding to $j=\frac32$. The gap  disappears again  for $\theta=\frac\pi2$, where we have the again a degenerate energy level corresponding to $n=2$ of the meta-Hydrogenoid spectrum $E^h_2$. If we increase the boundary condition parameter $\theta$  the gap reappears again. This time with the energy corresponding to $j=\frac32$ lower than the one corresponding to $j=-\frac32$. Finally, for $\theta=\pi$ the two energy levels become again degenerate at the level $n=2$ of $E^H_2$.\\

In the critical regime,  $\alpha^2=j^2$, as we have anticipated, the Hydrogenoid $E^{III}_n(0)$  and meta-Hydrogenoid $E^{III}_n(\pi)$ spectra do coincide   and are given by \eref{eig 1} with the only difference that for $\alpha^2=j^2$ and $j>0$,  $E_0=0$. Once more  this fact we can be understood in  a simpler way, just by looking at the corresponding boundary conditions. Analyzing how the spectrum $E^{III}_n(\theta)$ changes with $\theta$, we find that the correspondence in this case is
\begin{eqnarray}
\lim_{\theta\to0} E^{III}_n(\theta)&=&E^H_n,\label{S3S3}\\
\lim_{\theta\to\pi} E^{III}_n(\theta)&=&E^h_{n+1},
\end{eqnarray}
for $j>0$.
The spectrum is also periodic in this case, i.e. it is the same at $\theta$ and $\theta+\pi$, but the flow shifts the energy levels by one unit each time that we increase $\theta$ by $\pi$. In general, for fixed angular momentum and charge
we have
\begin{equation}
 E^{III}_n(\theta+k \pi)=E^{III}_{n+k} (\theta),\label{period}
\end{equation}
for any integer $k\in \Z$.
But even  in that case  the spectral flow  has  a monotonic character, i.e. $ E^{III}(\theta)< E^{III}(\theta')$ if 
$ \theta< \theta' $.
The inequalities between the Hydrogenoid and meta-Hydrogenoid energy levels \eref{rel eig2} 
 become the standard inequality
of the Hydrogenoid levels $E^H_n<E^H_{n+1}$ in this case.
The behaviour of the spectral flow recalls again the pumping mechanism of edge states in topological 
insulators \cite{Laughlin, Halperin}.

\begin{figure}[h!]
\centering{\includegraphics[width=13.5cm]{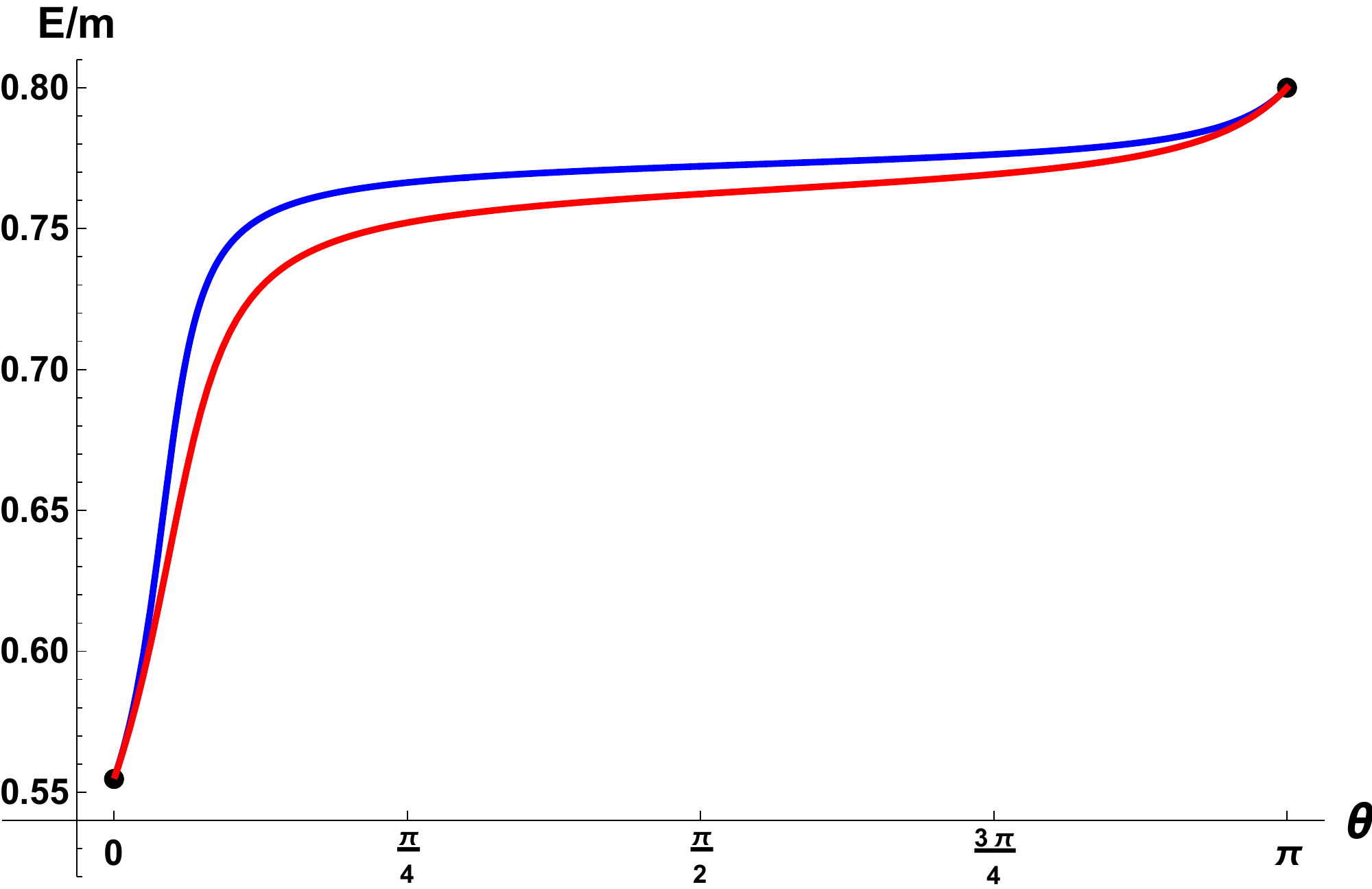}} 
\caption{Gap between the energies corresponding to $j=3/2$ (blue) and $j=-3/2$ (red) for $\alpha=3/2$ and 
$\Lambda=m/10$ in the critical regime.}
\label{fig gapcritical}
\end{figure}

The degeneracy between the bound energy levels with total angular momentum $j$ and $-j$  (for $n>0$) at $\theta=0$ and $\theta=\pi$ is again broken for intermediate values of $\theta\in(0,\pi)$ as shown in Figure \ref{fig gapcritical}. The level with negative angular momentum $-|j|$ has always lower energy than that of the corresponding level
with positive angular momentum $|j|$.

\begin{figure}[!h]
\centering{\includegraphics[width=14cm]{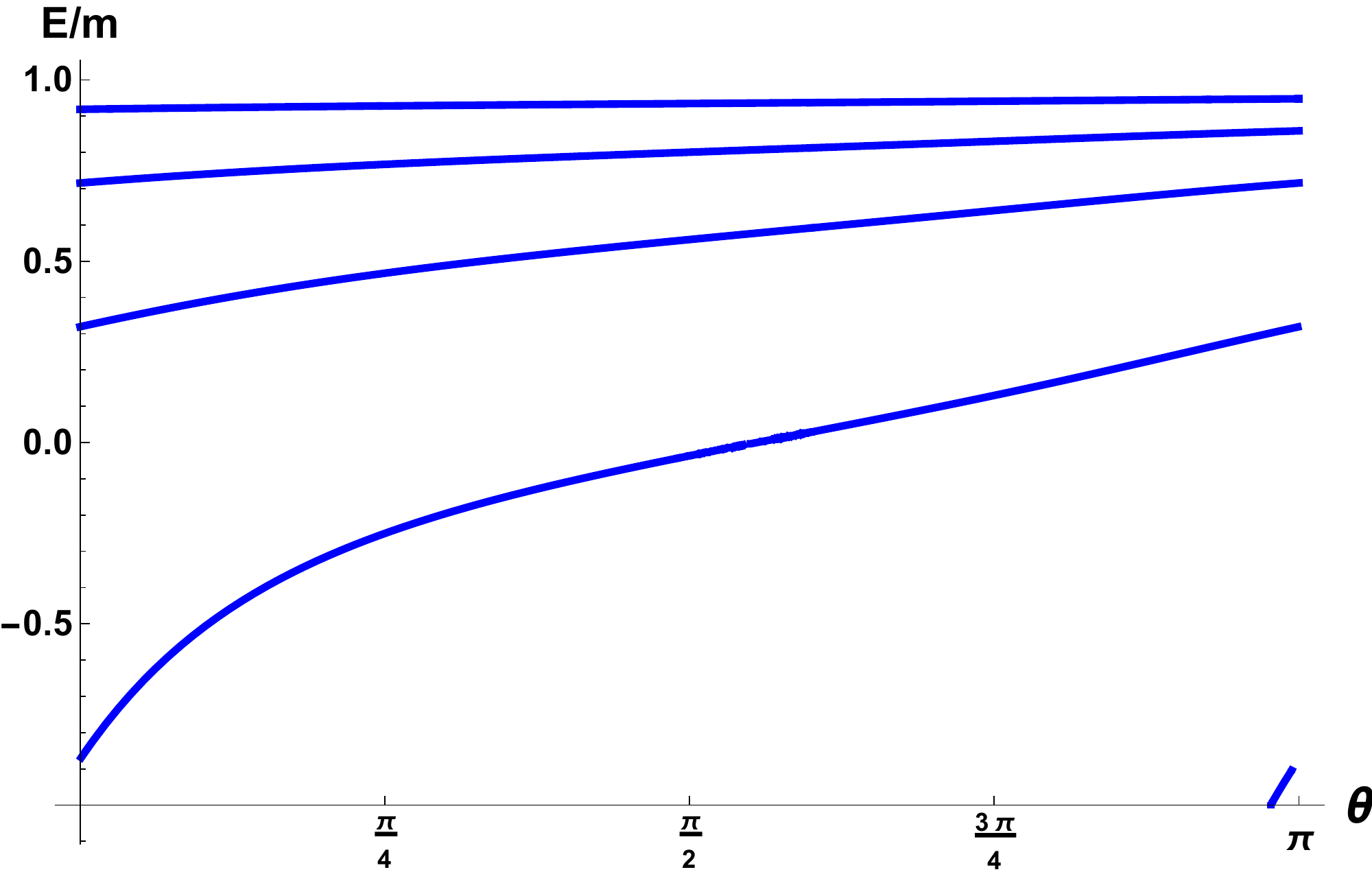}}
\caption{Spectral flow for $\alpha=1.55$, $\Lambda=m/10$ and $j=3/2$ in the supercritical regime. The lowest 
bound state energy level emerges from the negative continuum spectrum for a value of $\theta$ close to $\theta=\pi$.}
\label{fig super}
\end{figure}

 Let us now analyze the {\it supercritical charge regime} with  $ \alpha^2> j^2$. As anticipated, in this regime, the levels $E^H_n$ and $E^h_n$ do not belong to the spectrum of the Hamiltonian. In this case for any value of $\theta$ the energy spectrum
$E^{IV}_n(\theta)$ contains an infinity number of bound states that accumulate near  the mass gap continuum energy level $E=m$.  In Figure (\ref{fig super}) we plot the flow of some eigenvalues of the spectrum  $E^{IV}_n(\theta)$ when parameter 
$\theta$ flows  from $0$ to $\pi$. Notice that along that flow
one eigenvalue   pops up from the  Dirac sea continuum $E<-m$ at a particular value of the parameter $\theta$.

This is the only footprint of the instabilities pointed out in the supercritical regime, where the analytic expressions of Hydrogenoid and meta-Hydrogenoid energy levels become formally complex.  Notice that the same phenomenon occurs in the subcritical regime $ \alpha^2<j^2$. The  appearance of these instabilities is what inspired the Gribov  mechanism of quark confinement in QCD \cite{Gribov}--\cite{Gribov5} (see also \cite{AS}--\cite{AS5}).

\subsection{ Spectral flow and impurity charges}
\begin{figure}[h!]
  \centering{\includegraphics[width=14cm]{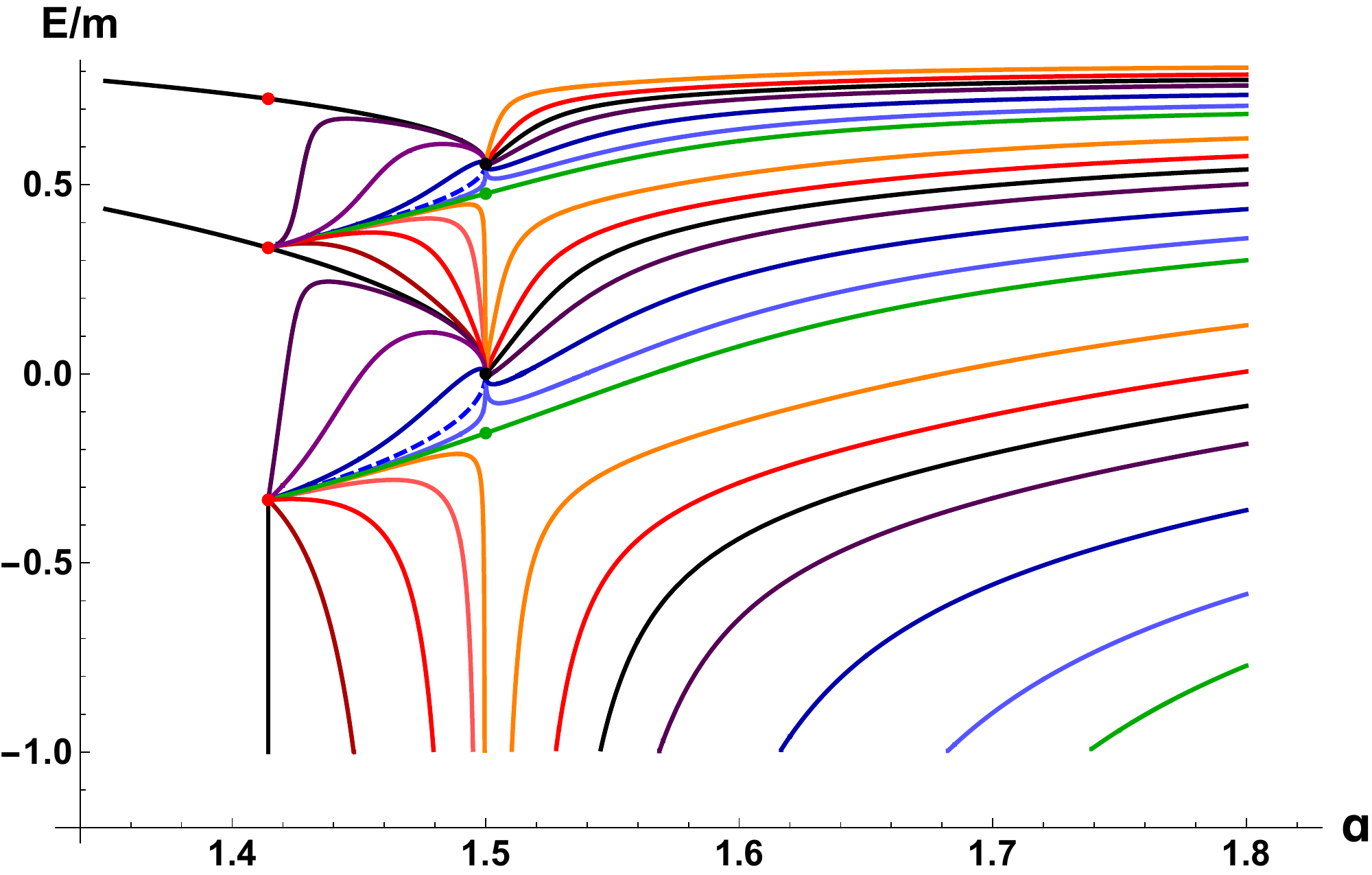}}
\caption{Flow  of the lowest energy levels with angular momentum   $j=3/2$  ($\Lambda=m/10$)
when the impurity charge crosses from  subcritical regime
to  supercritical regime at $\alpha=3/2$. The colors correspond to different choices of boundary conditions.
In the subcritical regime  
$\theta= 0$   (Hydrogen),
$\darkerred{\theta= 0.005\, \pi}$,
\red {$\theta=0.03 \, \pi$},
$\lighterred{\theta=0.1\, \pi}$,
$\orange{\theta=0.2\, \pi}$,
$\darkergreen{\theta=0.25\, \pi}$ (isolated),
$\lighterblue{\theta=0.3\, \pi}$,
$\blue{\theta=0.5\, \pi}$ (meta-Hydrogen),
$\darkerblue{\theta=0.9\, \pi}$,
$\purple{\theta=0.99\, \pi}$,
$\darkerpurple{\theta=0.999\, \pi}$; and in the 
supercritical regime 
$\theta=0$,
\red  {$\theta=0.1\, \pi$},
$\orange{\theta=0.25\, \pi}$,
$\darkergreen{\theta=0.5\, \pi}$ (isolated),
\blue {$\theta=0.6\, \pi$},
$\darkerblue{\theta=0.75\, \pi}$,
$\darkerpurple{\theta=0.9\, \pi}$.
}
\label{spectralflow}
\end{figure}

In order to analyze the transition from the subcritical regime to the supercritical regime we fix a suitable value of the parameter $\theta$ for $E^{II}_n(\theta)$, $E^{III}_n(\theta)$ and $E^{IV}_n(\theta)$. By increasing the value of  $ \alpha$ we can follow the flow of  each energy level from  the subcritical regime  to the critical regime  in an adiabatic continuous way. Notice, however,  
that for each  $0<\theta< \frac\pi4$  there is a bound state in the subcritical regime that merges to the continuum  for a special  value of $\alpha<j$
, and conversely, there is an infinity of bound states emerging from the continuum spectrum for $\alpha \gtrsim |j|$ in the supercritical regime for any $\theta$.
In any case  we have the following relations
\begin{equation}
\lim_{\alpha\rightarrow |j|_-}E^{II}_n(\theta)=E^{III}_n(0)=\lim_{\alpha\rightarrow |j|_+}E^{IV}_n(\theta');\label{continuity1}
\end{equation}
whenever $\theta\neq\frac\pi4$ and $\theta'\neq\frac\pi2$.
This means that, for any fixed values of $\theta$ ($\theta\neq \frac\pi4$ and $\theta'\neq \frac\pi2$),  $E^{II}_n(\theta)$ and $E^{IV}_n(\theta')$, converge to $E^{III}_n(0)$ as $\alpha\to \alpha=|j|$, pointing out the continuity of the flow of energy levels in the transition from the subcritical regime to the critical one (See  Figure \ref{spectralflow}).
In the exceptional cases we also have continuity in the  path crossing the transition border
\begin{equation}
\lim_{\alpha\rightarrow |j|_-}E^{II}_n(\textstyle \frac\pi4)=E^{III}_n(\textstyle \frac\pi2)=\lim_{\alpha\rightarrow |j|_+}E^{IV}_n(\textstyle \frac\pi2),\label{continuity1}
\end{equation}
provided we choose the suitable values for the parameter $\theta$ of the boundary condition in the different regimes.

The transition of the spectral flow from the subcritical to the supercritical regime is illustrated  in Figure 
\eref{spectralflow}. The flow shows the dependence  of  bound states on the impurity charge.
They are  also highly dependent on the boundary conditions of the different self adjoint extensions. For simplicity we consider only the angular momentum $j=\frac32$. 
For simplicity,  only the flow of lowest bound states of the infinite tower is displayed  for different values of  $\alpha$.
The flow of the higher
energy levels  is in fact very similar. In the region $0<\alpha<\sqrt{2}$, the operator is essentially self adjoint and the spectrum is that of an Hydrogenoid atom (black lines in Figure \eref{spectralflow}) that begin  at $E=m$ for $  \alpha=0$.  On the border of the subcritical region ($  \alpha=\sqrt{2}$) the smallest level (red point in Figure \eref{spectralflow}) is the ground state of the meta-Hydrogenoid spectrum, while  the other two levels  are doubly degenerated, because they include the $n$-level of the Hydrogenoid spectrum and  the $(n+1)$-level of the meta-Hydrogenoid spectrum. All the energy levels  of the different self adjoint extensions of $H$ start from one of these points.

At the critical coupling $  \alpha=\frac32$, we also have some special energy levels, which to some extent, are attractors or  repulsors of the other energy levels:  the black points correspond to the double degenerate 
Hydrogenoid and meta-Hydrogenoid spectra of $\theta=0$, whereas the green points correspond to the bound states of  the spectrum  of $H$ for $\theta=\pi/2$. 
The alternating black and green points  are, respectively, stable and unstable fixed points for the flow of energy levels. 
Each green point attracts only one energy level, that corresponds to the  self adjoint extension with  $\theta=\pi/4$ which are on  the green curve of  Figure \eref{spectralflow}.
 These flow lines are isolated and act as repulsive barriers  creating  bifurcations of the flow.  For $0\leq\theta<\pi/4$ the ground state merges into the continuum flowing to $-\infty$ and all  other $n+1$ levels flow into the $n$ black point, while for $\pi/4<\theta<\pi$ all  $n$ levels flow towards the $n$ black point ($n\geq0$). In the supercritical region, each green point is the starting point of  only one level (green curve) which is associated to a particular selfadjoint extension
  $\theta=\pi/2$.  whereas  black points  are the starting points of bound state energy levels for  all  other boundary conditions $\theta\neq\pi/2$. The green levels again are isolated and create a barrier for all the others. Notice that for any $\theta\neq\pi/2$ there are energy levels which  emerge from the continuum for large enough  values of the charge $\alpha$. In fact there is an infinity of them  if we consider higher excited bound states. In Figure \eref{spectralflow} we just displayed one of those levels emerging from the continuum for each boundary condition). 

\begin{figure}[h!]
\centering{\includegraphics[width=14cm]{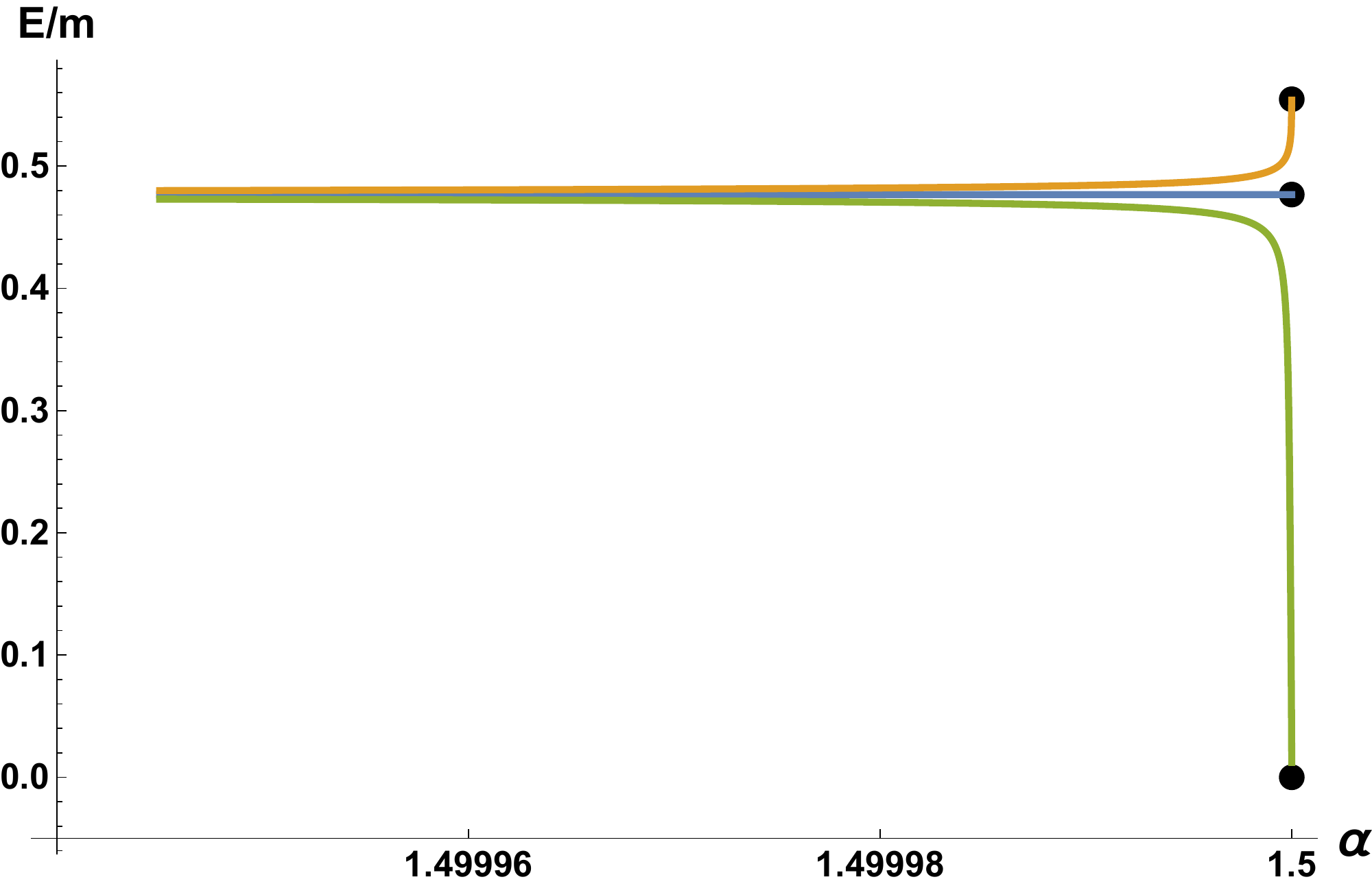}}
\caption{Instability properties of the flow of $E^{II}_n(\theta)$ for $\theta=\frac\pi4$ and $\theta_\pm=\frac\pi4\pm0.001$ (up/down) and   $j=3/2$ ($\Lambda=m/10$).}
\label{fig instab1}
\end{figure}

\begin{figure}[h!]
\centering{\includegraphics[width=14cm]{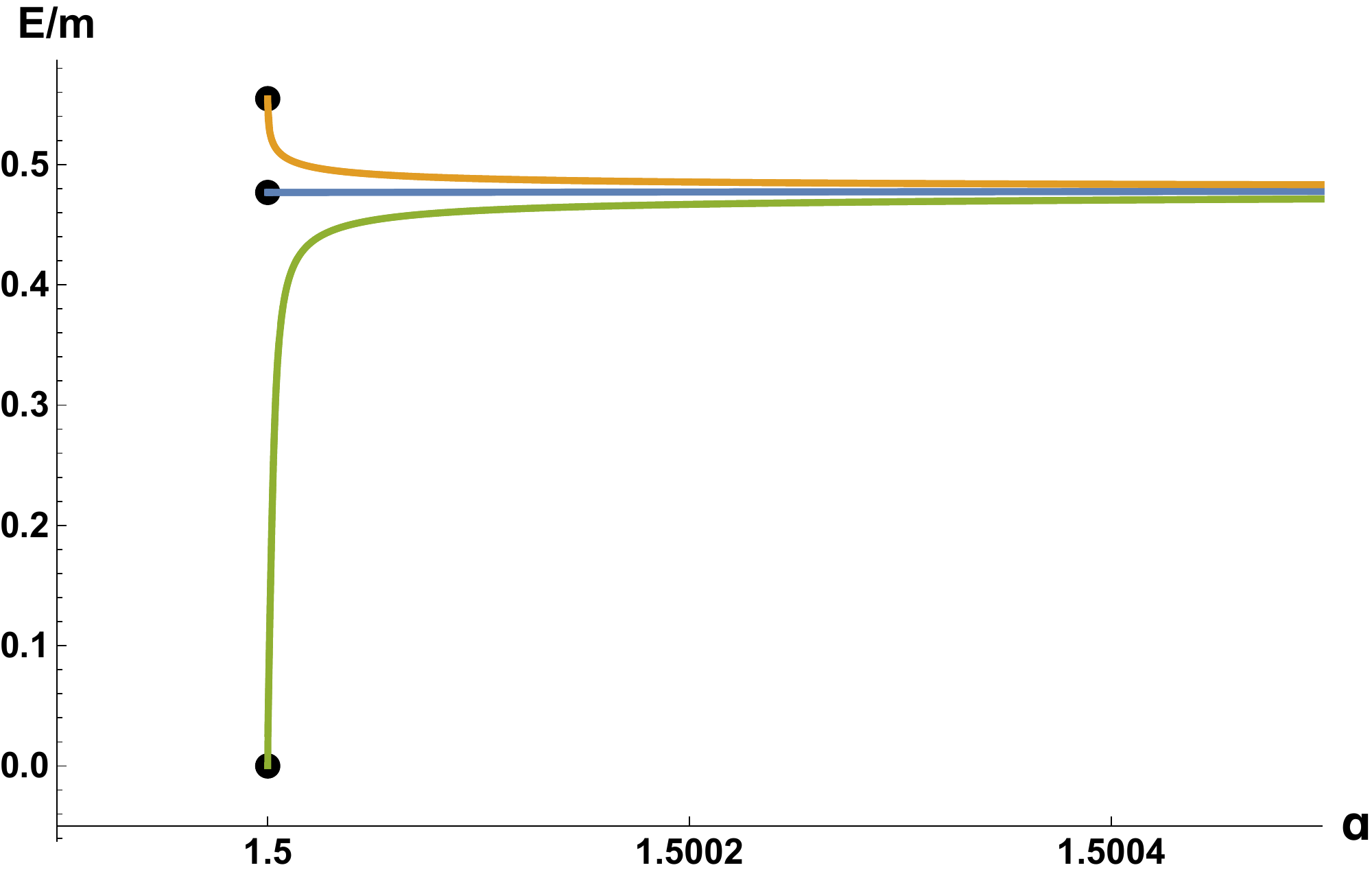}}
\caption{Instability properties of the flow of $E^{IV}_n(\theta)$ for $\theta=\frac\pi2$ and $\theta_\pm=\frac\pi2 \pm 0.005$ (up/down) and   $j=3/2$ ($\Lambda=m/10$).}
\label{fig instab2}
\end{figure}

In Figures (\ref{fig instab1}) and (\ref{fig instab2}) we show the instability of the {\it isolated} flow lines. The central flux lines  correspond, respectively to $\theta=\frac\pi4$ and $\theta=\frac\pi2$, while the others correspond to small perturbations of these lines, respectively $\theta=\frac\pi4\pm0.001$ and $\theta=\frac\pi2\pm0.005$. We can see how, when approaching to $\alpha^2=j^2$, the perturbed curves follow the {\it isolated} lines flow but eventually they are attracted by two different eigenvalues of  $E^{III}_n(\textstyle 0)$.

\section{Conclusions}

In summary, the above analysis in terms of boundary conditions shows that in graphene we have infinite set of self adjoint Dirac operator  for any  $\alpha>0$ (for $j=\frac12$) which are parameterized by an ultraviolet scale $\Lambda$ and an angle $\theta\in[0,\pi)$. In this sense the behaviour of impurities in graphene is different from that of Hydrogenoid atoms in QED. The lowest angular momentum states  in graphene are always in the subcritical regime unlike in the 3D Hydrogen atom, which requires the introduction of appropriate boundary conditions that depend on a UV scale $\lambda$ and a dimensionless parameter $\theta$. 

Unitarity is guaranteed for any value of the charge impurity $\alpha$. Even more, the parameters introduced by the boundary conditions $\Lambda, \theta$ that renormalize  the singular UV of the impurity induce remarkable observable effects.  The dependence on the choice of  boundary conditions at the singularity defines a  flow of  energy levels. The analysis of the flow of boundary levels displays interesting physical properties. Changes of the $\theta$ parameter which characterizes the self adjoint extension of the Hamiltonian can pump each Hydrogenoid  level into the next one after a recursive loop in the parameter space recalling the pumping mechanism of topological insulators. A Berry phase can be also associated to this process. All energy levels in the Hydrogenoid  spectrum, except the fundamental one, are degenerate, but the introduction of the parameter $\theta$ breaks  down this degeneracy. Moreover, it is possible to change, by adiabatical variations of $\alpha$, the energy levels from the subcritical to the supercritical regime in a continue way. Some bound states emerge (merge) from the continuum in this process.  
This is a consequence of the interesting properties of the RG flow for the subcritical and supercritical regime. Near the   critical charge the energy levels are attracted by the points of the spectra of the Hamiltonian at the critical charge $   \alpha^2=j^2$ and the particular value of the boundary conditions  $\theta=0$. The attracting Hamiltonian  corresponds to the Hydrogenoid  atom spectrum at the critical charge. Only few levels remain isolated in a unstable way.
This points out  that  the critical charge $ \alpha^2=j^2$ of the Hydrogenoid  case is not a singular case from the quantum physics viewpoint. The theory is well defined  below and above this critical charge in  the subcritical and supercritical regimes. The transition from the subcritical to the supercritical regime does not imply a 
critical change in the physical description of the system.

However, the apparent stability of the vacuum pointed out by the careful analysis of the boundary 
conditions of the Hamiltonian can not hide that the physical behaviour of  graphene is
quite special  in the supercritical phase. The fact that Hydrogenoid energy levels become complex
in the supercritical regime implies  the presence of resonances in the
spectral density of the scattering matrix in the positron (hole) channel. These resonances 
are also the root of  bound states levels which emerge from the continuum negative 
spectrum $E<-m$ (see Figure \ref{fig super}).

 In the supercritical regime there is an infinite number of
quasi-bound states embedded in the lower continuum $E<-m$ which are visible in the
spectral density. If they are not filled  when  cross the supercritical value, some normal
electrons will jump into these empty levels generating particle/hole pairs. The positive
charges will move to infinite and disappear whereas the negative charges remain localized near the impurity
giving rise to a screening of the impurity charge. We have assumed a positively charged impurity
but due to the CP invariance of the theory a similar phenomenon occurs for negative charged
impurities.

The phenomena described above are  reminiscent of what happens in Quantum Eletrodynamics \cite{Muller,Shabaev}.
The main difference  is that the value of the
critical charge in graphene is $\alpha=j$ whereas in QED is $Z=137$, which is very hard to
realize in Nature. The screening phenomenon due to supercritical pair creation has been
recently observed in graphene \cite{Wang, Wang0} and in QED a similar phenomenon might be also observed in
the heavy ions collisions  (see \cite{Rafelski} for an updated review). 
There is another remarkable difference between the two theories. In graphene
for any value of $\alpha>0$ the system is in a subcritical regime at least in the lowest angular momentum 
sector ($j=\half$) which requires always the choice of an extra parameter to fix
the boundary condition at the origin. However, in QED for $Z<137$, e.g. for the Hydrogen atom
the Hamiltonian is essentially selfadjoint in lowest angular momentum sector. 
Thus there is no need to fix the boundary condition at
the origin. In particular a potential $\delta$ like perturbation has no effect in the spectrum. In particular this means that
the relativistic interpretation of the Lamb effect cannot be understood in pure relativistic quantum mechanics and requires a full field theoretical analysis, unlike in the non-relativistic quantum mechanics approach.

The analysis of the energy spectrum in the gapless semi-metal regime  of graphene
can be carried out in a similar way. The boundary conditions are exactly the same  as in the massive case and,
thus,  the different physical regimes are the same. However, there is a fundamental difference, there are 
no electronic bound states, because the exponential decay at infinity disappears when $m\to 0$, although there are some   special  points of the continuous spectrum  that correspond to resonances which can be observed in scattering processes
\cite{Wang0, Wang}.

$$$$$$$$$$$$$$$$$$$$$$$$$$$$$$$$$$$$$$$$$$$$$$$$$$$$$$$$$$$$$$$$$$$$$$$$$$$$$$$$$$$$$$$$$$$$$$$$

\end{document}